\documentclass[review]{elsarticle}

\makeatletter
\def\ps@pprintTitle{%
 \def\@oddfoot{}%
 \let \@evenfoot\@oddfoot}
\makeatother

\usepackage{lineno,hyperref}
\usepackage{lineno,hyperref}
\usepackage{amsmath,amssymb}

\usepackage{tabularx}
\usepackage[ruled]{algorithm2e} 
\usepackage{algorithmic}

\usepackage{mathtools}
\usepackage{amsmath} 
\usepackage{amsthm}

\usepackage{amsfonts}
\usepackage{tikz}
\usepackage[utf8]{inputenc}

\newtheorem{theorem}{Theorem}
\newtheorem{lemma}{Lemma}
\newtheorem{corollary}{Corollary}

\theoremstyle{definition}

\begin{document}

\begin{frontmatter}

\title{A Note on Minimal Generating Sets for Semiflows}

\author{Gerard Memmi}
\address{LTCI, Telecom-Paris, Institut polytechnique de Paris}



\address{19 place Marguerite Perey F-91120 Palaiseau, France}

\begin{abstract}
\textit{In this note, we are interested in discussing characteristics of finite generating sets for $\mathcal{F^{+}}$, the set of all semiflows with non negative coordinates of a Petri Net. By systematically positioning these results over semirings such as $\mathbb{N}$ or $\mathbb{Q^+}$ then over a field such as $\mathbb{Q}$, we were able to complete the range of results about not only finite but optimal generating sets for a given family of semiflows.}
\end{abstract}


\end{frontmatter}


\section{Introduction}
The notion of generating set for semiflows is well known and support efficiently the handling of a specific class of invariants sometimes unduly called linear invariants in the literature. 
Several results have been published starting from their initial definition and structure \cite{M77} to 
a large array of applications especially to analyze Petri Nets \cite{Colom2003, DworLo16}.
Several algorithms were independently developed to compute a generating set of semiflows \cite{T81} or later in \cite{martinez1982, AM82}. 
All of them can be considered as variations of Farkas or Fourier algorithms related to integer linear programming and convex geometry 
(see \cite{ColomS89} for a comparative study or \cite{Schrijver87} for the underlying mathematical theory).

Identifying the set of minimal supports is critical for developing effective formal analysis of Petri Nets and proving complex behavioral properties (including parameters). 
This provides its full significance to the three decomposition theorems in section \ref{sec: 3theorems} that are slightly revisited an are at the core of this note motivated by the need to regroup a number of results in the literature closely related to them. 

In this short note, we are interested in discussing characteristics of finite generating sets for $\mathcal{F^{+}}$, the set of all semiflows with non negative coordinates of a Petri Net.
The notion of minimal generating sets for semiflows described in section \ref{subsec: mgs} can be found in \cite{M77}, then by Colom, Silva, and Teruel in \cite{STC1998} p319 and later by the same authors in \cite{GV03}, p 68 or more recently, in \cite{CMPW09}.
By systematically comparing and positioning these results over semirings such as $\mathbb{N}$ or $\mathbb{Q^+}$ then over a field such as $\mathbb{Q}$, we were able to complete as accurately as possible the range of results about finite minimal generating sets for a given family of semiflows. 
To this effect, simple examples and counterexamples were provided to illustrate our results and show a few differences between minimal or canonical semiflows. 
However, theorem \ref{th: lgs=mgs} is new to the best of our knowledge and prove the coincidence between a least generating set and a minimal generating set. 
Uniqueness of particular generating sets is presented section \ref{subsec: uniqueness}.

In section \ref{sec: example-summary}, a long example illustrates how liveness of Petri Nets can be proven with semiflows, and in particular how the usage of  semiflows of minimal support bring simplicity to the analysis. 
A table is summarizing main results, in particular showing differences when considering $\mathbb{N}$, $\mathbb{Q^+}$, or $\mathbb{Q}$ 
\footnote{these differences are at the source of few inaccurate statements found in the literature \cite{ColomS89, CMPW09}} before concluding the note.

\subsection{Petri Nets and semiflows}
A \textit{Petri Net} is a tuple $PN = <P, T, Pre, Post>$ where $P$ is a finite set of \textit{places}, $T$ a finite set of \textit{transitions} such that $P \cap T = \text{\O}$. 
A transition $t$ of $T$ is defined by its $Pre(\cdot,t)$ and $Post(\cdot,t)$ \textit{conditions} \footnote{We use here the usual notation: $Pre(\cdot,t)(p) = Pre(p,t)$ and  $Post(\cdot,t)(p) = Post(p,t)$} :
$Pre: P \times T \rightarrow \mathbb{N}$ is a function providing a weight for pairs ordered from places to transitions, $Post: P \times T \rightarrow \mathbb{N}$ is a function providing a weight for pairs ordered from transitions to places. 
$d$ will denote the number of places: $d = |P|$.

Extensive definitions, properties, and case studies can be found in particular in \cite{BR82, GV03}.
 
\textit{Semiflows} can be defined as solutions of the following homogeneous system of $|T|$ diophantine equations: 
\begin{equation}
\label{eq:inv-semiflow}
    f^T Post(\cdot,t) = f^T Pre(\cdot,t), \ \ \forall t \in T
\end{equation}
where $x^T y$ denotes the scalar product of the two vectors $x$ and $y$. 

Equation \ref{eq:inv-semiflow} allows to directly deduce the following invariant from a non null semmiflow $f$:

for any marking $M$ reachable from a given marking $M_0$, we have : 
\begin{equation}
\label{eq:inv}
    f^T M = f^T M_0
\end{equation}
In the sequel, we will consider uniquely the set $\mathcal{F^{+}}$ of semiflows with non negative coordinates (semiflows with negative coordinates can also be defined for instance in \cite{ColomS89}; however, they are not the object of this note). Then, $\mathcal{F^{+}}$ can be defined by:

$\mathcal{F^{+}} = \{f \in \mathbb{N}^{d} \ | \ \forall t \in T, f^T Post(\cdot,t) = f^T Pre(\cdot,t)\}$.

We abusively use the same symbol ‘0' to denote $(0,...,0)^T$ of $\mathbb{N}{^n}\ \forall n \in \mathbb{N}$. 
The \textit{support} of a semiflow $f$ is denoted by $\left \| f \right \|$ and is defined by:

$\left \| f \right \| = \{x\in P\ |\ f(x) \neq 0\}$.

We will use the usual the componentwise 
partial order, in which

$(x_1, x_2, \dots, x_d )^T \leq ( y_1, y_2, \dots, y_d )^T$ if and only if $x_i \leq y_i\ \forall i \in \{1, \dots, d\}$.

\section{Generating sets and minimality}

\subsection{Basic definitions and results}
\label{subsec : basic}
A subset $\mathcal{G}$ of $\mathcal{F^{+}}$ is a \textit{generating set over a set $\mathbb{S}$} if and only if $\forall f \in \mathcal{F}$ we have $f = \sum_{g_i \in \mathcal{G}} \alpha_ig_i $ where $\alpha_i \in \mathbb{S}$, $g_i \in \mathcal{G}$, and $\mathbb{S} \in \{\mathbb{N}, \mathbb{Q^{+}}, \mathbb{Q}\}$ where $\mathbb{Q^{+}}$ denotes the set of non negative rational numbers.
Since $\mathbb{N} \subset \mathbb{Q^{+}} \subset \mathbb{Q}$ \footnote{ where $\subset$ denotes the strict inclusion between sets}, a generating set over $\mathbb{N}$ is also a generating set over $\mathbb{Q^{+}}$ and a generating set over $\mathbb{Q^{+}}$ is also a generating set over $\mathbb{Q}$. The reverse is indeed not true and in our opinion is one source of inaccurate propositions that can be found in the literature.

The notion of generating set is strongly related with algebraic concepts especially when the generating set is finite. 
Let's consider $\mathcal{G}$ a finite generating set such that $\mathcal{G} \{g_1,...g_q\}$, the following definitions can be recalled:

If $\mathcal{G}$ is a generating set over $\mathbb{N}$ then $\mathcal{C(G)} = \{ f \in \mathbb{N}{^d}\ |\ f = \sum_{i=1}^{i=q} \alpha_ig_i\}$ where $\alpha_i \in \mathbb{N}$ is called a \textit{semigroup} and $\mathcal{F^{+}} = \mathcal{C(G)}$.

If $\mathcal{G}$ is a generating set over $\mathbb{Q}^+$ then $Cone(\mathcal{G}) = \{ f \in (\mathbb{{Q}^+}){^d}\ |\ f = \sum_{i=1}^{i=q} \alpha_ig_i\}$ where $\alpha_i \in \mathbb{Q}^+$ is called a convex polyhedral cone and $\mathcal{F^{+}} = Cone(\mathcal{G}) \cap \mathbb{N}{^d}$. 
It is interesting to recall a result from \cite{Lasserre89} stating that $\mathcal{F^{+}} \neq \{0\}$ if and only if $Cone(\mathcal{G}) \neq \{0\}$.

If $\mathcal{G}$ is a generating set over $\mathbb{Q}$ then $VS(\mathcal{G}) = \{ f \in (\mathbb{{Q}^+}){^d}\ |\ f = \sum_{i=1}^{i=q} \alpha_ig_i\}$ where $\alpha_i \in \mathbb{Q}$ is called a vector space and $\mathcal{F^{+}} = VS(\mathcal{G}) \cap \mathbb{N}{^d}$. We can extract from $\mathcal{G}$ a basis of $VS(\mathcal{G})$ which also is a generating set of $\mathcal{F^{+}}$ over $\mathbb{Q}$ (see for instance \cite{Lan02} p. 85).

The fact that there exists a finite generating set over $\mathbb{N}$ is non trivial. This result has been proven by Gordan circa 1885 then Dickson circa 1913. Here, we directly rewrite Gordan's lemma \cite{AB86} by adapting it to our notations:

\begin{lemma} (\textbf{Gordan circa 1885})
\label{lem : Gordan}
Let  $\mathcal{F^+}$ be the set of non-negative integer solutions of equation \ref{eq:inv-semiflow}. Then, there exists a finite generating set of vectors in $\mathcal{F^+}$, such that every element of $\mathcal{F^+}$ is a linear combination of these vectors with non-negative integer coefficients.
\end{lemma}

In the sequel, we will not come back on the question of the existence of a finite generating set. Being shown for $\mathbb{N}$, it is necessarily true for $\mathbb{Q^+}$ and $\mathbb{Q}$. 

\subsection{Minimal supports and minimal semiflows}
Several definitions of the notion of minimal semiflow have been proposed in \cite{STC1998} p.319, in \cite{GV03} p.68, \cite{Kruck86}, \cite{CMPW09}, or in \cite{M78, M83}. It can be confusing to look into these in details. Rather, we propose to consider only two basic notions in order theory: minimality of support with respect to set inclusion and minimality of semiflow with respect to the componentwise partial order on $\mathbb{N}{^d}$ since the various definitions we found in the literature as well as the results of this note can be described in terms of these sole two classic notions.

A non empty support $I$ of a semiflow $f$ is \textit{minimal} with respect to the set inclusion if and only if $\nexists \ g \in \mathcal{F}-\{0\}$ such that $\left \| g \right \| \subset I$. 

Since $P$ is finite, the set $\mathcal{MS}$ of minimal supports in P is a \textit{Sperner family} (i.e. a family of subsets such that none of them contains another one) and we can apply Sperner's theorem \cite{Sper28} over $m=|\mathcal{MS}|$ which is stating that:
\begin{equation}
\label{eq: Sperner}
    m \leq  \binom{d}{\left \lfloor d/2 \right \rfloor}
\end{equation}.
This provides us with a general upper bound for the number of minimal supports in a given Petri Net (this result is already given in \cite{M83, STC1998}). To the best of our knowledge, this bound can be reached only by three families of degenerated Petri Nets containing isolated elements: only one, two, or three places and as many isolated transitions as desired. We conjecture that this bound cannot be reached for Petri Nets with more than three places and should be refined on a case-by-case basis by exploiting connectivity between places and transitions as hinted in the example section \ref{sec: example-summary}.

A non null semiflow $f$ is \textit{minimal} with respect to $\leq$ if and only if $\nexists \ g \in \mathcal{F^{+}}-\{0,f\}$ such that $g \leq f$.

In other words, we cannot decompose a minimal semiflow as the sum of another semiflow and a non null non negative vector. This remark provides us with a first hint on how fundamental the notions of minimality are regarding semiflows decompostion and to finding out how semiflows can be generated by as small as possible a finite subset of semiflows each of them with as small a support as possible. One reason for looking after minimality of semiflows and supports is to reduce the complexity of the analysis of various behavioral properties; given first that the number of minimal semiflows can be great second, that considering a basis may not capture constraints deduced from minimal support as easily (see example section \ref{sec: example-summary}).

\section{Three decomposition theorems}
\label{sec: 3theorems}
Generating sets can be characterized thanks to a set of three decomposition theorems that can be found in  \cite{M78} with their proofs. Here, theorem \ref{th: over N} slightly extended is provided with a new proof by using Gordan's lemma \ref{lem : Gordan}.
Theorem \ref{th: min support} is extended to include considerations regarding $\mathbb{Q^+}$ and $\mathbb{Q}$. 
\begin{theorem}
\label{th: over N}

If a semiflow is minimal then it belongs to any generating set over $\mathbb{N}$.

The set of minimal semiflows of $\mathcal{F^{+}}$ is a finite generating set over $\mathbb{N}$.
\end{theorem}
Let's consider a semiflow $f \in \mathcal{F^{+}}-\{0\}$ such that $\exists f_1,...,f_k \in \mathcal{F^{+}}-\{0,f\}$ and $a_1,...,a_k \in \mathbb{N}$ 
such that $f = \sum_{i=1}^{i=k}a_if_i$. 
Since $f \neq 0$ and all coefficients $a_i$ are in $\mathbb{N}$, $\exists j \leq k$ such that $a_j > 0$. 
Therefore, $a_jf_j \leq f$, since $f_j \neq f$, we have $f_j<f$ so $f$ is not minimal.
Hence, if a semiflow is minimal then it has to belong to every generating sets over $\mathbb{N}$.

Applying Gordan's lemma, there exists $\mathcal{G}$ a finite generating set. Since any minimal semiflow is in $\mathcal{G}$, the subset of all minimal semiflows is included in $\mathcal{G}$ and therefore finite. Let $\mathcal{E} = \{e_1,...e_n\}$ be this subset. 

For any semiflow $f \in \mathcal{F^+}$, we always can define $r \in \mathcal{F^+}$ and a set of $n$ non negative integers $\{k_1,...k_n\}$ such that:

i) $r = f- \sum_{j=1}^{j=n}k_je_j$ 

ii)  $\forall i \leq n,
(f- \sum_{j=1}^{j=i}k_je_j) \in \mathcal{F^+}$
and $(f - \sum_{j=1}^{j=i}k_je_j -e_i) \notin \mathcal{F^+}$

If $r \neq 0$ then we would have $r \in \mathcal{F^+}-\{0\}$ and by construction, $\nexists e_i \in \mathcal{E}$ such that $e_i \leq r$ which would mean that $r$ is minimal and contradict the fact that $\mathcal{E}$ includes all minimal semiflows.
Therefore, $r=0$ and any semiflows can be decomposed as a linear combinations of minimal semiflows and $\mathcal{E}$ is a generating set.\footnote{if $\mathcal{E}$ was infinite the construction could still be used (since the decreasing sequence is bounded by 0 and $\mathbb{N}$ is nowhere dense) and we would have:
\newline
$ \lim \limits_{n \to \infty} f- \sum_{j=1}^{j=n}k_je_j = 0$ with the same definition of the coefficients $k_j$ as in ii).}
\hfill
$\square$

\begin{figure}[ht]
\centering
\includegraphics[width=0.35
\textwidth]{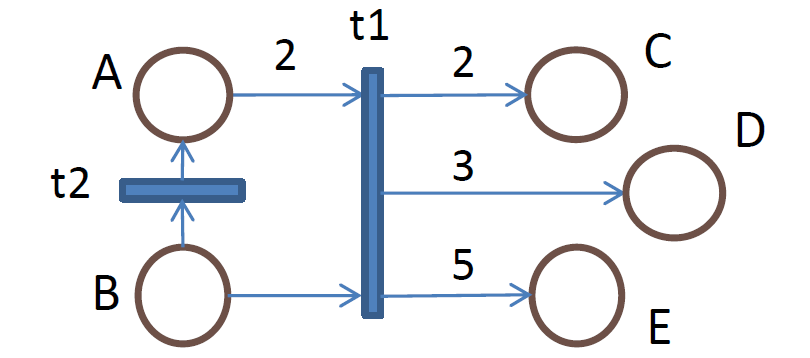}
\caption{$f_1^T=(3,3,2,0,1),\ f_2^T=(4,4,1,0,2),\ g_1^T=(2,2,3,0,0),\ g_2^T=(1,1,0,1,0),\ g_3^T=(5,5,0,0,3)$ are five canonical and minimal semiflows. $\left \| f_1 \right \|$ or $\left \| f_2 \right \|$ are not minimal. $f_1$ and $f_2$ are linear combinations of $g_1, g_2, g_3$: $f_1= \frac{1}{3}(2g_1+g_3)$ and $f_2= \frac{1}{3}(g _1+2g_3)$. Therefore, the decomposition of a semiflow on $\{f_1, f_2, g_1, g_2, g_3\}$ is not unique. Moreover, $\mathcal{G}_1=\{g_1, g_2, g_3\}$ constitutes a generating set over $\mathbb{Q^+}$ or over $\mathbb{Q}$.} 
\label{STC3}
\end{figure}

Let's point out that since $\mathcal{E}$ is not necessarily a basis, the exhibited decomposition is not necessarily unique and depends on the order in which the minimal semiflows are considered as shown in figure \ref{STC3}.

\begin{figure}[ht]
\centering
\includegraphics[width=0.3\textwidth]{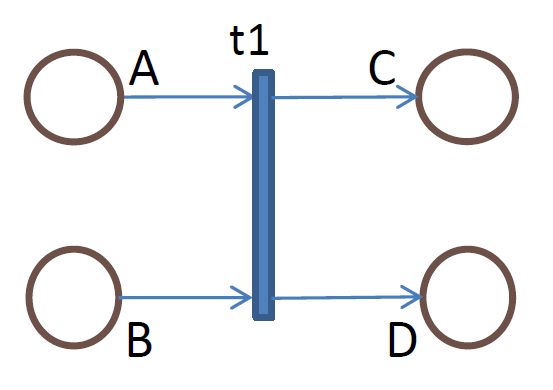}
\caption{$f^T=(1,1,1,1),\ g_1^T=(0,1,1,0),\ g_2^T=(0,1,0,1),\ g_3^T=(1,0,1,0),\ g_4^T=(1,0,0,1)$ are five canonical semiflows. However, $f$ is not minimal and $\left \| f \right \|$ is not minimal. $\mathcal{G}_2=\{g_1,g_2,g_3,g_4\}$ is a unique generating set over $\mathbb{N}$ and $f= g_1+g_4=g_2+g_3$ has exactly two different decompositions in $\mathcal{G}_2$. $\mathcal{G}_3 = \{g_1, g_2, g_3\}$ is a generating set over $\mathbb{Q}$.}
\label{STC2}
\end{figure}

However, a minimal semiflow does not necessarily belong to a generating set over $\mathbb{Q^{+}}$ or $\mathbb{Q}$. This is illustrated in figure \ref{STC3} where $\mathcal{G}_1$ does not include $f_1$ which is minimal or in figure \ref{STC2} where $\mathcal{G}_3$ does not include $g_4$ which is minimal of minimal support. 

\begin{theorem}
\label{th: min support}
If $I$ is a minimal support then 

i) there exists a unique minimal semiflow $f$ such that $I = \left \| f \right \|$ and $\forall g \in \mathcal{F^{+}}$ such that $\left \| g \right \| = I, \exists k \in \mathbb{N}$ such that $g = kf$,

ii) any non null semiflow $g$ such that $\left \| g \right \| = I$ constitutes a generating set over $\mathbb{Q^{+}}$ or $\mathbb{Q}$ for $\mathcal{F}_I^+ = \left\{ g \in \mathcal{F^+} |\ \left\| g\right\|= I\right\}$.
\end{theorem}
In other words, $\{f\}$ is a unique generating set over $\mathbb{N}$ for $\mathcal{F}_I^+ =\{g \in \mathcal{F^{+}} \ | \ \left \| g \right \| = I\}$. 
However, this uniqueness property is indeed lost in $\mathbb{Q^{+}}$ or in $\mathbb{Q}$ since any element of $\mathcal{F}_I^+$ is a generating set over $\mathbb{Q^{+}}$ or $\mathbb{Q}$.

From Sperner's theorem, we know that any support $I$ of a semiflow contains a finite number $m$ of minimal supports of semiflows. The following theorem states that these $m$ supports first, cover $I$, second, provide a generating set directly deduced from these m supports.
\begin{theorem} (decomposition)
\label{th: decomp}
Any support $I$ of semiflows is covered by the finite subset $\{I_1, I_2, \dots, I_m\}$ of minimal supports of semiflows included in $I$:

$I = \bigcup_{i=1}^{i=m} I_i$

Moreover,

$\forall f \in \mathcal{F^{+}}$ such that $\left \| f \right \| = I$, we have $f=\sum_{i=1}^{i=m} \alpha_ig_i$ where $\alpha_i \in \mathbb{Q^{+}}$ and the semiflows $g_i$ are such that $\left \| g_i \right \| = I_i$
\end{theorem}

A sketch of proof of theorem \ref{th: decomp} can be found in \cite{BR82}, a complete proof in \cite{M78}.


 \section{Canonical semiflows}

In \cite{ColomS89} or in \cite{GV03} p68, a semiflow is said \textit{canonical} if and only if the gcd of its non null coordinates is equal to one (In \cite{CMPW09}, such a semiflow is said to be \textit{scaled back}).

Minimal semiflows and canonical semiflows are two different notions. The following lemma and theorem help at comparing them.

\begin{lemma}
\label{lem : canonical-minimal}
If a semiflow is minimal then it is canonical.

If a semiflow is canonical and its support is minimal then it is minimal.
\end{lemma}

The first point is quite evident: if $f$ is not canonical its gcd $k$ is such that $k>1$ so $\exists g \in \mathcal{F^{+}}$ such that $f= kg$ and $f$ would not be minimal.

The second point is a direct application of theorem \ref{th: min support}.
\hfill
$\square$

However, canonical semiflows are not necessarily minimal semiflows; and minimal semiflows have not necessarily a minimal support. 
In the example of figure \ref{STC3}, $f_1$ and $f_2$ are canonical and minimal but their support is not minimal.
Any semiflow $f= ag_1 + g_2 + bg_3$ where $a, b \in \mathbb{N}$ and $a + b > 0$ is canonical and obviously not minimal. As $a$ and $b$ are as  great as wanted, this is showing that the number of canonical semiflows can be infinite. Moreover, we can observe in this example that the support of this sequence of infinite canonical semiflows is not minimal. This is hinting on the following theorem

\begin{theorem}
\label{th: canonical}
Given a support $I$, $C_I\ = |\{f canonical\ semiflow \ | \ \left \| f \right \| = I \}|$,

If $I$ is a minimal support then $C_I = 1$ else $C_I$ is infinite.

If $C_I = 1$ then $I$ is minimal
\end{theorem}

From theorem \ref{th: min support}, there is a unique minimal semiflow having a given minimal support. From lemma \ref{lem : canonical-minimal}, a minimal semiflow is canonical. Hence, If $I$ is minimal then $C_I = 1$.

If $I$ is not minimal then $\exists\ e,\ f \in \mathcal{F}^+$ such that $ \ \left \| e \right \| \subset \ \left \| f \right \| = I$.
We can build an infinite sequence of semiflows $f_i, i \in \mathbb{N}$ such that 
$f_i = \alpha_i (f + k_ie)$ where $1/\alpha_i$ is the gcd of the non null coordinates of $f_i$, and $k_i \in \mathbb{N}$. 
$\forall i \in \mathbb{N}, f_i$ is canonical by construction. 
Let's consider $i,j$ such that $f_i=f_j$; then $\alpha_i (f + k_ie) = \alpha_j (f + k_j)$. 
This leads to: $ (\alpha_i - \alpha_j)f = (k_i - k_j)e$. 
However, since $ \ \left \| e \right \| \subset \ \left \| f \right \|$, we must have $\alpha_i = \alpha_j$ and $k_i = k_j$. 
Hence, we built an infinite sequence of canonical semiflows based upon the infinite sequence of non negative integers.

If $C_I = 1$ then let's $f$ be the unique canonical semiflow of support $I$. 
Let's consider $g \in \mathcal{F}^+$ such that  
$ \ \left \| g \right \| \subseteq I$. 
With the same construction as before, we can build a canonical semiflow $h= \alpha (f + kg)$ where $1/\alpha_i$ is the gcd of the non null coordinates of $h$, and $k \in \mathbb{N}$. We have $ \ \left \| h \right \| = I$ and $C_I = 1$, therefore, $h=\alpha (f + kg)=f$. Then, $g=((1-\alpha)/k\alpha)f$ which means that any semiflow of support included in $I$ is a multiple of $f$. Hence, $I$ is minimal.
\hfill
$\square$

The fact that the number of canonical semiflows can be infinite was already pointed out in \cite{Kruck86, ColomS89, CMPW09}. The fact that this number is infinite only when the considered support is non minimal as described in theorem \ref{th: canonical} is new to the best of our knowledge.

 \section{Minimal generating sets, least generating sets, and fundamental sets}
\label{subsec: mgs}


Minimal generating sets have been defined over $\mathbb{N}$ in \cite{M78}, over $\mathbb{Q^+}$ in \cite{M78, M83}, and least generating sets over $\mathbb{Q}$ in \cite{ColomS89, GV03}.
Similarly to the notion of generating set defined in section \ref{subsec : basic}, we slightly extend their definition to hold over a set $\mathbb{S} \in \{\mathbb{N}, \ \mathbb{Q^+}, \mathbb{Q}\}$. 


From \cite{M83} p 39, a \textit{minimal generating set} over $\mathbb{S}$ is a generating set that does not strictly include any generating set.

From \cite{ColomS89} p 82, or in \cite{GV03} p 68, a \textit{least generating set of semiflows} "is made up of the least number of elements to generate any semiflow" over $\mathbb{S}$ \footnote{More precisely, the least generating set is defined over $\mathbb{Q}$ in \cite{ColomS89, GV03} and over $\mathbb{N}$ in \cite{CMPW09}.}
In other words, $\mathcal{G}$ is a least generating set if and only if it does not exist a generating set $\mathcal{H}$ such that $\left| \mathcal{H} \right| \leq \left| \mathcal{G} \right|$. 

A minimal generating set is defined with respect to set inclusion while a least generating set is defined with respect to its cardinality. In the case of generating sets of semiflows, theorem \ref{th: lgs=mgs} hereunder is a new result stating that these two different notions are in fact equivalent over $\mathbb{S}$.

\begin{lemma}
\label{lemma: lgs=mgs}
If $\mathcal{G}$ is a generating set over $\mathbb{Q^+}$ or $\mathbb{N}$, $I$ a minimal support, then $\exists g \in  \mathcal{G}$ such that $I= \left\|g \right\|$.
\end{lemma}
We consider $e$ a semiflow of minimal support, $\mathcal{G} = \{g_1,...g_k\}$, a generating set over $\mathbb{Q^{+}}$ or $\mathbb{N}$. Then, $e = \sum_{i=1}^{i=k}\alpha_ig_i$. 
All the coefficients are non negative and $e \neq 0$ then, $\exists j \leq k$ such that $\alpha_j > 0$ and $e \geq \alpha_jg_j$. Since $\left \| e \right \|$ is minimal, $\left \| e \right \| = \left \| g_j \right \|$. 
\hfill
$\square$

This lemma states that any generating set over $\mathbb{Q^+}$ or $\mathbb{N}$ contains at least one semiflow per minimal support. Indeed, this point is not true over $\mathbb{Q}$. In figure \ref{STC2}, $\mathcal{G}_2$ is a minimal generating set over $\mathbb{Q}^+$ and $\{g_2, g_3, g_4\} \subset \mathcal{G}_2$ is a generating set over $\mathbb{Q}$ since $g_1= g_2 + g_3 -g_4$ is of minimal support but generated over $\mathbb{Q}$ (since one coefficient is negative) by the other minimal semiflows of minimal support.

\begin{theorem}
\label{th: lgs=mgs}

If $\mathcal{G}$ is a generating set over $\mathbb{S}$, where $\mathbb{S} \in \{\mathbb{N}, \mathbb{Q^{+}}, \mathbb{Q}\}$ then the two properties are equivalent:

$\mathcal{G}$ is a minimal generating set,

$\mathcal{G}$ is a least generating set. 
\end{theorem}

First, the fact that a least generating set be a minimal generating set is straightforward and true in general. 

Let's consider $\mathcal{G}$, a minimal generating set over $\mathbb{N}$. By applying theorem \ref{th: over N}, we directly conclude that $\mathcal{G}$ is the set of minimal semiflows and is a least generating set.

Let's consider $\mathcal{G}$, a minimal generating set over $\mathbb{Q}^{+}$. Then, the lemma \ref{lemma: lgs=mgs} can apply stating that $\mathcal{G}$ includes $\mathcal{G'}$ a family of exactly one semiflow for each minimal support. From theorem \ref{th: decomp}, we draw that $\mathcal{G'}$ is a generating set. $\mathcal{G}$ is minimal then $\mathcal{G} = \mathcal{G'}$. This being true for any minimal generating set, $\mathcal{G}$ is also a least generating set.

Let's consider $\mathcal{G}$, a minimal generating set over $\mathbb{Q}$. From $\mathcal{G}$ we can extract a subset $\mathcal{B}$ of linearly independent semiflows (see \cite{Lan02} p. 85 for basic results on vector spaces). Then, $\mathcal{B}$ is a least and minimal generating set over $\mathbb{Q}$. 
\hfill
$\square$

Theorem \ref{th: canonical} states that there is exactly one canonical semiflow for each minimal support. This particularity characterizes the notion of fundamental set.


In \cite{STC1998} p 319, the set of all canonical semiflows of minimal support is called  \textit{fundamental set}.

\begin{corollary} (fundamental set) 
A fundamental set is a generating set over $\mathbb{Q^{+}}$ or $\mathbb{Q}$ but not necessarily over $\mathbb{N}$.

A fundamental set over $\mathbb{Q^{+}}$ is a minimal generating set but not necessarily over $\mathbb{Q}$. 
\end{corollary}

The first point of this corollary is a direct consequence of theorem \ref{th: decomp} and lemma \ref{lem : canonical-minimal}: we conclude that a fundamental set is one possible generating set over $\mathbb{Q^{+}}$ and therefore over $\mathbb{Q}$.

the second point is directly deduced from the first point and lemma \ref{lemma: lgs=mgs}.

The last parts of the two points of the corollary are illustrated by the two following counterexamples:

In figure \ref{STC3}, $\mathcal{G}_1=\{g_1, g_2, g_3\}$ is a fundamental set which is not a generating set over $\mathbb{N}$ since $f_1$ or $f_2$ are minimal and cannot be decomposed as a linear combination of elements of $\mathcal{G}_1$ over $\mathbb{N}$ \footnote{To this regard, the statement p.143-147 of \cite{CMPW09} must be rewritten.}.

A fundamental set over $\mathbb{Q}$ is not necessarily a minimal generating set: in figure \ref{STC2}, $\mathcal{G}_2$ is a fundamental set and is not a minimal generating set.
\hfill
$\square$

Also, the fact that a semiflow $f$ belongs to a least generating set denoted by $lgs$ does not provide it with a lot a properties \footnote{the properties 2.2 p. 82 of \cite{ColomS89}, 5.2.5 p.68 of \cite{GV03} must be rewritten by taking the following statements into account}:

If $f \in lgs$ over $\mathbb{N}$ then $f$ is minimal but not necessarily canonical or of minimal support.

If $f \in lgs$ over $\mathbb{Q}^+$ then $f$ has a minimal support but is not necessarily canonical or minimal.

$f \in lgs$ over $\mathbb{Q}$ then $f$ is not necessarily minimal, not necessarily canonical or of minimal support.

\subsection{About uniqueness}
\label{subsec: uniqueness}
\begin{corollary}(uniqueness)
The set of minimal semiflows is the unique minimal generating set over $\mathbb{N}$.

If $\mathcal{G}$ is a least generating set over $\mathbb{Q}^+$ then for any minimal support $I$ of semiflows, $\exists g \in \mathcal{G}$ unique such that $I = \left \| g \right \|$.

The fundamental set is the unique minimal generating set of minimal semiflows over $\mathbb{Q}^{+}$. In a given Petri Net, there exists only one fundamental set.
\end{corollary}
The first point can be directly deduced from theorem \ref{th: over N}, the second and third ones are directly deduced from theorems \ref{th: min support}; \ref{th: decomp} and lemma \ref{lem : canonical-minimal}.
\hfill
$\square$

The third point of this corollary can be found in \cite{STC1998}.

However, a minimal generating set over $\mathbb{Q}^+$ or $\mathbb{Q}$ is not unique even among minimal semiflows of minimal support. 
In the example of figure \ref{STC2}, 

$\{k_1g_1, k_2g_2, k_3g_3, k_4g_4\}$ where $k_i \in \mathbb{N}$ constitutes a family of minimal generating sets over $\mathbb{Q}^+$. 
Also, $ \mathcal{G}_3 = \{g_1,g_2,g_3\}$ and $\{g_2,g_3,g_4\}$ are two minimal generating sets over $\mathbb{Q}$.

\section{Results summary}
\label{sec: summary}
\begin{figure}[ht]
\centering
\includegraphics[width=0.9\textwidth]{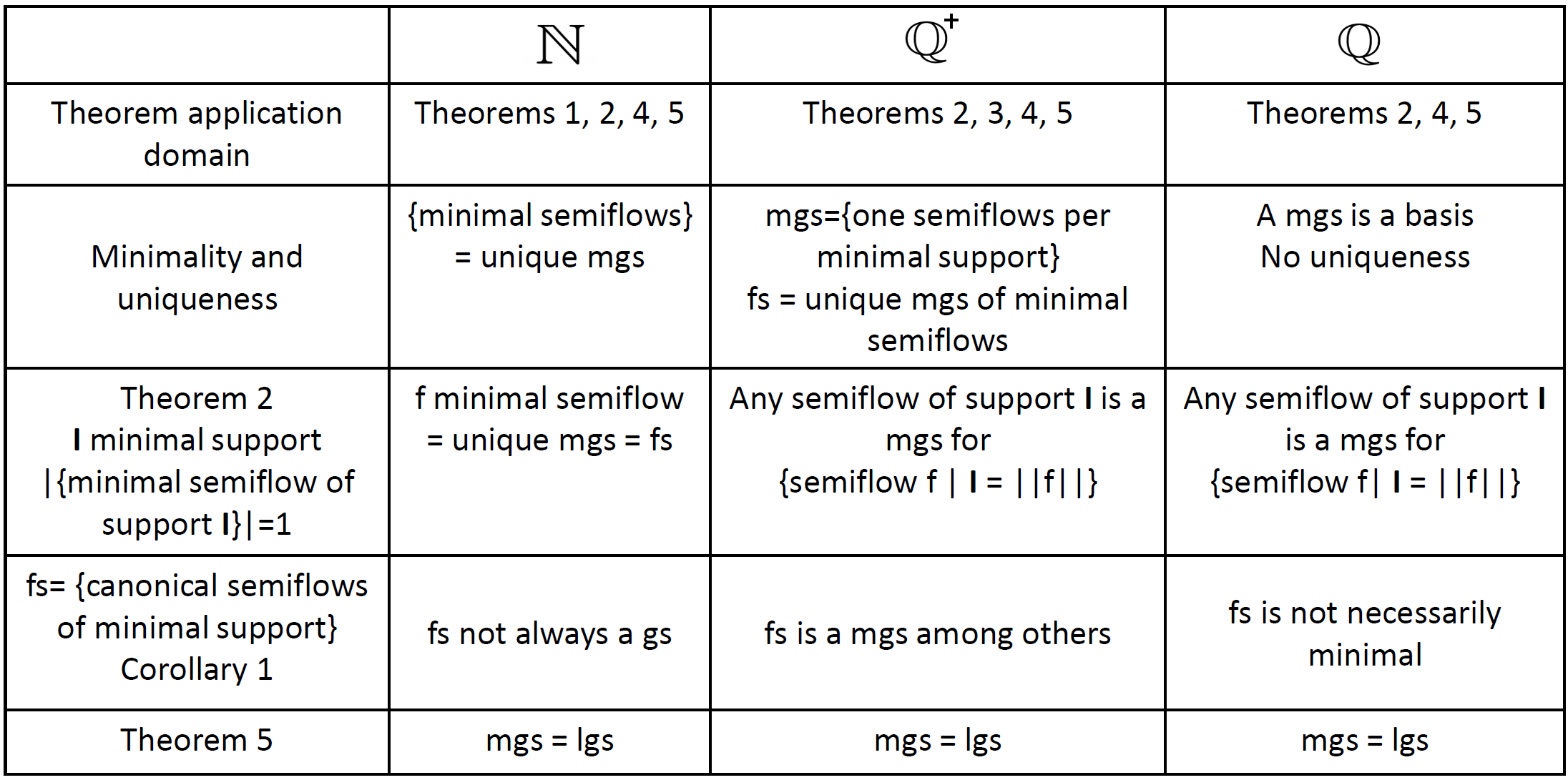}
\caption{gs, fs, mgs, lgs denote generating set, fundamental set, minimal generating set, least generating set respectively.}
\label{table-gs}
\end{figure}

We have seen through several counterexamples that no result must be taken for granted and that any proposition must be carefully proven. 
The table of figure \ref{table-gs} is summarizing the main results of this note.

\section{Reasoning with invariants, an example}
\label{sec: example-summary}
The example described figure \ref{Mame} is a simplified version of a Petri Net published in \cite{MeMa81} representing two subscribers (in idle state in places $LA$ and $A$) who will have a conversation (places $CLA$ and $CA$) then hang up (transitions $t_3, t_4, t_5$ for LA and $t_8$ for A) before going back to their idle states. During these operations, they exchange messages $PU,\ R$ from $LA$ to $A$ and $S,\ F$ from $A$ to $LA$.

First, let us notice that inequation \ref{eq: Sperner} gives us the following bound for $m$, the number of supports:
$ m \leq \binom{9}{\left \lfloor 9/2 \right \rfloor}=126 $. However, it is easy to notice that the transitions $t_1, t_2, t_8, t_9$ have only one input and one output which means that any support including such an input also includes the corresponding output to verify the equations associated to $t_1, t_2, t_8, t_9$. We can reduce the bound and have: $m=\binom{5}{\left \lfloor 5/2 \right \rfloor}=10$. It would be possible to improve again this bound by reasoning on the other transitions.

\subsection{Using two different generating sets for the same analysis}
We then want to prove that the initial marking $M_0$ such that $M_0(LA)=M_0(A)=1, \ M_0(p)=0$ for any other place, is a home state (i.e. a marking such that whatever is the evolution of the Petri Net, it is always possible to reach it back) from which it is easy to deduce that the Petri Net is live.

\begin{figure}[ht]
\centering
\includegraphics[width=0.9\textwidth]{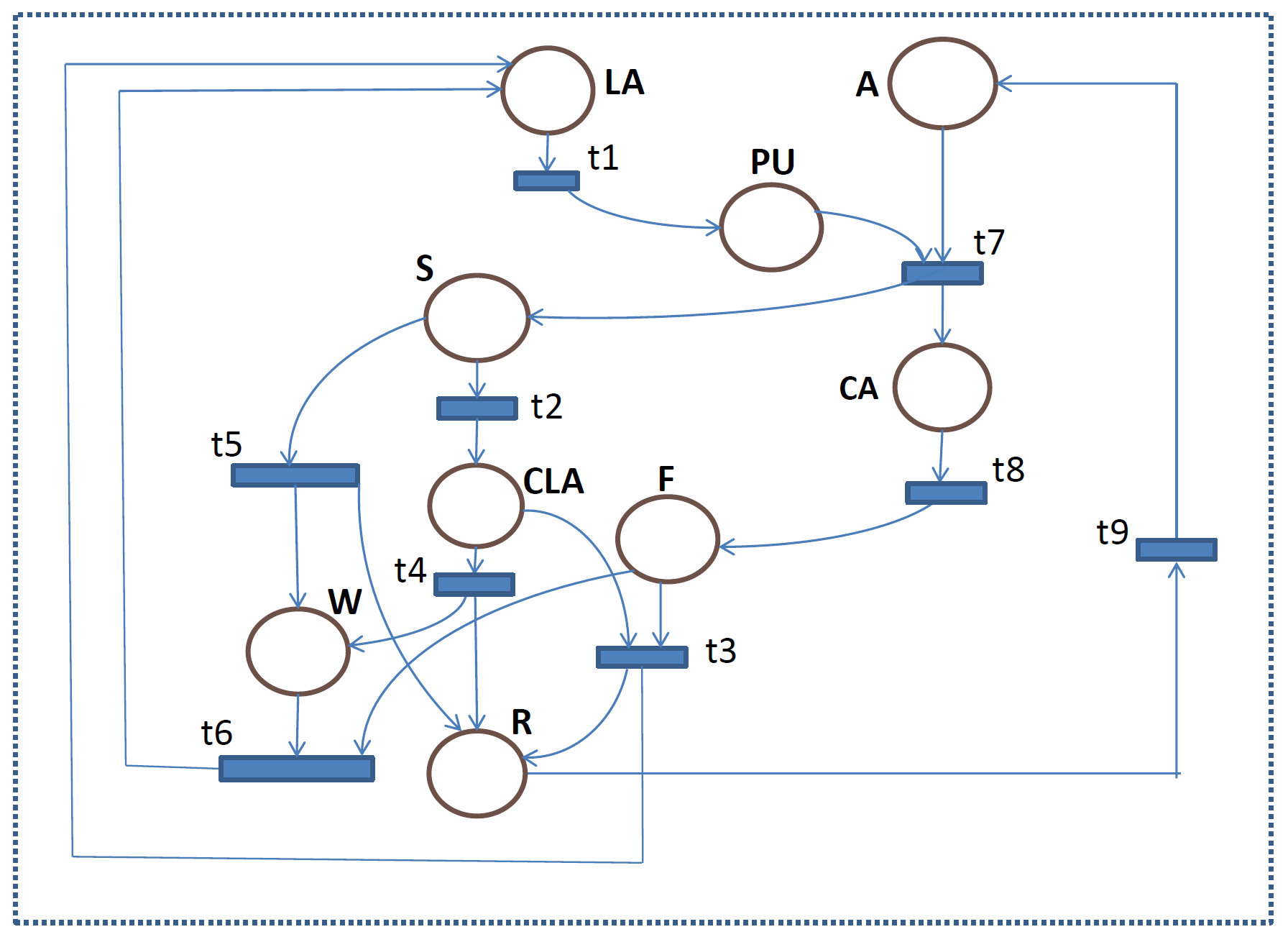}
\caption{This Petri Net has exactly three minimal semiflows of minimal support $f_1, f_2, f_3$ such that: 
\newline
$f_1(LA)=f_1(CLA)=f_1(W)=f_1(PU)=f_1(S)=\ 1, \ f_1(p)=\ 0$ for any other place,
\newline
$f_2(LA)=f_2(PU)=f_2(F)=f_2(CA)=\ 1,\ f_2(p)=\ 0$ for any other place,
\newline
$f_3(CLA)=f_3(S)=f_3(R)=f_3(A)= \ 1,\ f_3(p)=\ 0$ for any other place.
}
\label{Mame}
\end{figure}

We are going to prove this important property starting from two different generating sets. The first one is a minimal generating set over $\mathbb{Q^+}$ formed with $\mathcal{GB}_1=\{f_1,f_2,f_3\}$ the set of minimal semiflows of minimal support defined figure \ref{Mame}; the second one is a minimal generating set over $\mathbb{Q}$ defined by $\mathcal{GB}_2=\{f_1,\  g=f_2+f_3,\ h=f_1+f_3\}$.

From $\mathcal{GB}_1$, three invariants can be drawn:
$f_1^TM=1$, $f_2^TM=1$, $f_3^TM=1$ for any reachable marking from $M_0$.
In other words, there is exactly one token in the support of any semiflow of $\mathcal{GB}_1$.

To prove that $M_0$ is a home state, we start to prove that

$\mathcal{HS}=\{M |\ M(LA)=1\}$ is a home space (i.e. a set of marking such that whatever is the evolution of the Petri Net, it is always possible to reach one element of the set). If we consider the invariant deduced from $f_2$, we know that any reachable marking M is in one of the four following cases:

i) $M(LA)=1$, then $M \in \mathcal{HS}$,

ii) $M(F)=1$, since $f_2^TM=1$, we have $M(LA)=M(PU)=M(CA)=0$.
From $f_1^TM=1$, we have three sub-cases:

ii-1) $M(CLA)=1$, $t_3$ can occur from $M$ and we can reach $M' \in \mathcal{HS}$ with $M'(LA)=1$ 

ii-2) $M(W)=1$, $t_6$ can occur from $M$ and we can reach $M' \in \mathcal{HS}$ with $M'(LA)=1$

ii-3) $M(S)=1$, $t_2$ can occur from $M$ then we are in the sub-case ii-1).

iii) $M(CA)=1$, $t_8$ can occur from $M$ then we are in the case ii).

iv) $M(PU)=1$, then considering the invariants deduced from $f_1$ and $f_2$, we have: 
$M(LA)=M(CLA)=M(W)=M(F)=M(S)=M(CA)=0$

From $f_3^TM=1$, we have two sub-cases:

iv-1) $M(A)=1$, $t_7$ can occur from $M$ then we are in the case iii).

iv-2) $M(R)=1$, the sequence $t_9t_7$ can occur then we are in the case iii) again. 

From these 4 cases and 5 sub-cases, we directly deduce that $\mathcal{HS}$ is a home space from where it is easy to conclude that $M_0$ is a home state and the Petri Net is live.

Let us now develop the same method of proof with the second generating set $\mathcal{GB}_2$. This time, we consider the invariant deduced from $f_1$, we know that any reachable marking M is in one of the five following cases:

i) $M(LA)=1$, then $M \in \mathcal{HS}$, 

ii) $M(W)=1$, then considering the invariants deduced from $f_1$, we have: 

$M(LA)=M(CLA)=M(PU)=M(S)=0$

From $h^TM=2$, we have three possible sub-cases:

ii-1) $M(A)=1$, this time, $g^TM=2$ can be used to deduce two similar sub-sub-cases:

ii-1-$\alpha$) $M(F)=1$, $t_6$ can occur from $M$ and we can reach $M' \in \mathcal{HS}$ with $M'(LA)=1$

ii-1-$\beta$) $M(CA)=1$, the sequence $t_8t_6$ can occur with the same result as in sub-sub-case ii-1-$\alpha$).

ii-2) $M(R)=1$, $t_9$ can occur from $M$ and we are in sub-case ii-1).

ii-3) $M(W)=2$, this last sub-case does not satisfy the invariant deduced from $f_1$ and must be discarded. 

iii) $MCLA)=1$, then considering the invariants deduced from $f_1$ and $h$, we have: 
$M(LA)=M(W)=M(PU)=M(S)=M(R)=M(A)=0$

Similarly as in case ii-1), we use $g^TM=2$ and have two sub-cases:

iii-1) $M(F)=1$, $t_3$ can occur from $M$ and we can reach $M' \in \mathcal{HS}$ with $M'(LA)=1$

iii-2) $M(CA)=1$, the sequence $t_8t_3$ can occur with the same result as in sub-case iii-1).

iv) $M(S)=1$, $t_2$ can occur from $M$ and we are back in the iii).

v) $M(PU)=1$,then considering the invariants deduced from $f_1$, we have: 

$M(LA)=M(W)=M(CLA)=M(S)=0$,

we then use $g^TM=2$ and have two sub-cases:

v-1) $M(A)=1$, $t_7$ can occur from $M$ and we can reach the case iii),

v-2) $M(R)=1$, $t_9$ can occur from $M$ and we can reach the case v-1).

Since we started we the same amount of information, we fortunately reach the same conclusion however the second time with 5 more complex cases and 7 sub-cases and 2 sub-sub cases. We conclude that the smaller the support are, the more effective the analysis will be since cases and sub cases depend on the number of elements of each considered support. Of course, this is to balance with the complexity of computing minimal supports.

The same proof scheme could be used the same Petri Net enriched with parameters to model $x$ subscribers of type $LA$ and $y$ of type $A$.

\section{Conclusion}

By considering $\mathbb{N}$, then $\mathbb{Q^+}$, then $\mathbb{Q}$, the size of a minimal generating set decreases as expected since more and more possibilities to combine semiflows are provided. More interestingly, if $m$ is the number of minimal supports in a Petri Net then:

a minimal generating set over $\mathbb{N}$ is finite and has at least $m$ elements,

a minimal generating set over $\mathbb{Q^+}$ has exactly $m$ elements,

a minimal generating set over $\mathbb{Q}$ has at most $m$ elements.

Looking at canonical semiflows appears to us less significant. 

We have seen through the example of section \ref{sec: example-summary} how gross the Sperner's bound is as well as a first idea to improve it. The same example provides us with a reason to consider non negative semiflows of minimal support. There exist other ones described in \cite{Colom2003}.


We have seen that the results in \cite{GV03} p68 need to be rephrased and rather consider the previous publication of the same authors \cite{STC1998, CMPW09}. 


At last, we believe that these results may  be extended along two dimensions. From a mathematical point of view, the relation with integer linear programming has been published many times particularly in \cite{ColomS89}, however, a private communication with D. Madore suggests to look at the notion of toric varieties and saturated semigroups. From a Petri Net theory point of view, it remains to apply them to a variety of Petri Nets (such as colored Petri Nets).

\bibliography{mybibfile}

\end{document}